\documentclass[sigconf]{acmart}

\usepackage{booktabs} 

\usepackage{graphicx}
\usepackage{epstopdf}
\usepackage{amssymb}
\usepackage{amsmath}
\usepackage{subcaption}
\usepackage{epsfig}
\usepackage{color}
\usepackage{enumitem}
\usepackage{float}
\usepackage{soul}
\usepackage{wrapfig}

\usepackage{tikz}
\usetikzlibrary{shapes.geometric, arrows, fit, positioning, decorations.markings}
\usepackage{url}

\setcopyright{rightsretained}

\newtheorem{Remark 1}{Remark}
\newtheorem{Remark 2}[Remark 1]{Remark}
\newtheorem{Remark 3}[Remark 1]{Remark}
\newtheorem{Remark 4}[Remark 1]{Remark}
\newtheorem{Remark 5}[Remark 1]{Remark}
\newtheorem{Remark 6}[Remark 1]{Remark}
\newtheorem{Remark 7}[Remark 1]{Remark}

\newtheorem{Theorem 1}{Theorem}
\newtheorem{Theorem 2}[Theorem 1]{Theorem}
\newtheorem{Theorem 3}[Theorem 1]{Theorem}
\newtheorem{Theorem 4}[Theorem 1]{Theorem}
\newtheorem{Theorem 5}[Theorem 1]{Theorem}
\newtheorem{Theorem 6}[Theorem 1]{Theorem}
\newtheorem{Theorem 7}[Theorem 1]{Theorem}
\newtheorem{Theorem 8}[Theorem 1]{Theorem}
\newtheorem{Theorem 9}[Theorem 1]{Theorem}
\newtheorem{Theorem 10}[Theorem 1]{Theorem}

\begin{document}

\copyrightyear{2017}
\acmYear{2017}
\setcopyright{acmlicensed}
\acmConference{CPS-SPC'17}{November 3, 2017}{Dallas, TX,
USA}\acmPrice{15.00}\acmDOI{10.1145/3140241.3140243}
\acmISBN{978-1-4503-5394-6/17/11}
\fancyhead{}
\settopmatter{printacmref=false, printfolios=false}

\title{Secure and Privacy-Preserving Average Consensus}

\author{Minghao Ruan, Muaz Ahmad, and Yongqiang Wang}
\affiliation{%
  \institution{Department of Electrical and Computer Engineering}
  \streetaddress{P.O. Box 1212}
  \city{Clemson University} 
}
\email{yongqiw@clemson.edu}


\begin{abstract}
Average consensus is fundamental for distributed systems since it underpins key functionalities of such systems ranging from distributed information fusion, decision-making, to decentralized control. In order to reach an agreement, existing average consensus algorithms require each agent to exchange explicit state information with its neighbors. This leads to the disclosure of private state information, which is undesirable in cases where privacy is of concern. In this paper, we propose a novel approach that enables secure and privacy-preserving average consensus in a decentralized architecture in the absence of any trusted third-parties. By leveraging homomorphic cryptography, our approach can guarantee consensus to the exact value in a deterministic manner. The proposed approach is light-weight in computation and communication, and applicable to time-varying interaction topology cases. A hardware implementation is presented to demonstrate the capability of our approach.
\end{abstract}

%
%

\keywords{Average consensus, privacy}

\maketitle

\section{Introduction}
As a building block of distributed computing, average consensus has been an active research topic in computer science and optimization for decades \cite{Morris74, Lynch96}. In recent years, with the advances of wireless communications and embedded systems, particularly the advent of wireless sensor networks and the Internet-of-Things, average consensus is finding increased applications in fields as diverse as automatic control, signal processing, social sciences, robotics, and optimization \cite{Olfati-Saber2007}.  

Conventional average consensus approaches employ the explicit exchange of state values among neighboring nodes to reach agreement on the average computation. Such an explicit exchange of state information has two disadvantages. First, it results in breaches of the privacy of participating nodes who want to keep their data confidential. For example, a group of individuals using average consensus to compute a common opinion may want keep secret their individual personal opinions \cite{citeulike84}. Another example is power systems where multiple generators want to reach agreement on cost while keeping their individual generation information private \cite{Zhang11}. Secondly, storing or exchanging information in the plaintext form (without encryption) is vulnerable to attackers which try to steal information by hacking into the communication links or even the nodes. With the increased number of reported attack events and the growing awareness of security, keeping data encrypted in storage and communications has become the norm in many applications, particularly many real-time sensing and control systems such as the power systems and wireless sensor networks.

To address the pressing need for privacy and security, recently, several relevant average consensus approaches have been proposed. Most of these approaches use the idea of obfuscation to mask the true state values by adding carefully-designed noise on the state. Such approaches usually exploit tools such as mean-square statistics \cite{Mo17} or ``differential privacy'' which is heavily used for database privacy in computer science \cite{Erfan15, Huang15, Manitara13}. Although enhances privacy, such noise-based obfuscation also unavoidably affects the performance of average consensus, either directly preventing converging to the true value, or making convergence only achievable in the statistical mean-square sense. Furthermore, these approaches normally rely on the assumption of time-invariant interaction graph, which is difficult to satisfy in many practical applications where the interaction patterns may vary due to node mobility or fading communication channels. 

Neither can the above noise-based approaches protect nodes from attackers which try to steal information by hacking into the nodes or the communication channels. To improve resilience to such attacks, a common approach is to employ cryptography. However, it is worth noting that although cryptography based approaches can easily provide privacy and security when a trusted third-party is available, like in the multi-party computation \cite{Lagendijk13}, their extension to completely \textit{decentralized} average consensus without any \textit{trusted} third-parties are extremely difficult due to the difficulties in the decentralized management of keys. In fact, in the only reported result incorporating cryptography into decentralized average consensus \cite{Lazzeretti14}, privacy is obtained by paying the price of depriving participating nodes from access to the final consensus value, although partial information such as a binary decision is still retrievable for participating nodes. 

In this paper, we propose a homomorphic cryptography based approach that can guarantee privacy and security in decentralized average consensus even in the presence of a time-varying interaction graph. Different from existing noise-based privacy-preserving approaches which can only achieve average consensus in the statistic case, our approach can guarantee convergence to the \textit{exact} average value in a \textit{deterministic} manner. Unlike the existing cryptography based average consensus approach in \cite{Lazzeretti14}, this approach allows every participating nodes to access the exact final value. Furthermore, the approach is completely decentralized and light-weight in computation, which makes it easily applicable to resource restricted systems. 

The outline of this paper is as follows. Section \ref{sec:background} reviews the protocol used for average consensus problem and the homomorphic cryptography, particularly the Paillier cryptosystem. Our encrypted protocol is introduced in Section \ref{sec:proto}. In Section \ref{sec:analysis} we provide a proof of convergence and bounds of critical parameter, followed by a systematic discussion of  privacy guarantees as well as security enforcement mechanisms in Section \ref{sec:security}. Implementation issues and a physical implementation example are presented in Section \ref{sec:impl}. The conclusion is drawn in Section \ref{sec:conclusion}.

\section{Background}\label{sec:background}
In this section we briefly review the average consensus problem and
the homomorphic encryption.

\subsection{Average Consensus}
We follow the same convention as in \cite{Olfati-Saber2007} where a
network of $M$ nodes is represented by a graph
${G=(V,\,E,\,\mathbf{A})}$ with node set ${V}=\{v_1,\,v_2,\,\cdots,
v_M\}$, edge set ${E}\subset {V}\times {V}$, and a weighted
adjacency matrix $\mathbf{A}=[a_{ij}]$ which satisfies $a_{ij}>0$ if
$(v_i,v_j)\in E$ and 0 otherwise. The set of neighbors of a node
$v_i$ is denoted as
\begin{equation}
{N}_i = \left\{v_j \in {V}| (v_i,v_j)\in {E}\right\}
\end{equation}
Throughout this paper we assume that the graph is undirected and
connected. Therefore, $\mathbf{A}$ is symmetric
\begin{equation}
\label{eq:aij}
a^{(t)}_{ij} = a^{(t)}_{ji}>0\quad \forall (v_i,v_j)\in {E}
\end{equation}
Note that   the superscript $t$ denotes that the weights are
time-varying. Sometimes we drop $t$ for the sake of notation
simplicity, but it is worth noting that all discussions in the paper
are always applicable under time-varying weights. To achieve average
consensus, namely converging of all states $x_i(t)$
$(i=1,2,\cdots,M)$ to the average of   initial values, i.e.,
$\frac{\sum_{i=1}^M x_i(0)}{M}$, one commonly-used update rule for
the continuous-time (CT) domain is
\begin{equation}
\label{eq:ct}
\dot{x}_i(t) = \sum_{v_j\in {N}_i} a^{(t)}_{ij}\cdot(x_j(t)- x_i(t))
\end{equation}
The counterpart for discrete time (DT) is
\begin{equation}
\label{eq:dt}
x_i[k+1] = x_i[k] + \varepsilon\sum_{v_j\in N_i} a^{(k)}_{ij}\cdot(x_j[k]- x_i[k])
\end{equation}
where $\varepsilon$ is a constant step size residing in the range
$(0, 1]$.

%

\subsection{Homomorphic Encryption}
Our method to protect privacy and security is to encrypt  the state.
To this end, we briefly introduce a cryptosystem, more specifically
the public-key cryptosystem which is applicable in open and dynamic
networks without the assist of any trusted third party for key
management. Many popular cryptosystems  such as RSA
\cite{Rivest1978}, ElGamal \cite{ElGamal1985}, and Paillier
\cite{Paillier1999} are public-key cryptosystems. In this paper we
focus on the Pailler cryptosystem which provides the following basic
functions:
\begin{itemize}
	\item Key generation:
	\begin{enumerate}
		\item Choose two large prime numbers $p$ and $q$ of equal bit-length and compute $n=pq$.
		\item Let $g=n+1$.
		\item Let $\lambda= \phi(n)=(p-1)(q-1)$ where $\phi(\cdot)$ is the Euler's totient function.
		\item Let $\mu=\phi(n)^{-1}\;\text{mod}\;n$ which is the modular multiplicative inverse of $\phi(n)$.
		\item The public key $k_p$ is then $(n,g)$.
		\item The private key $k_s$ is then $(\lambda,\mu)$.
	\end{enumerate}
	\item Encryption ($c=\mathcal{E}(m)$):\\
	Recall the definitions of $\mathbb{Z}_n= \{z|z \in \mathbb{Z}, 0 \leq z < n\}$ and $\mathbb{Z}^*_n = \{z|z \in \mathbb{Z}, 0 \leq z < n, \text{gcd}(z, n) = 1\}$ where $\text{gcd}(a,b)$ is the greatest common divisor of $a$ and $b$.
	\begin{enumerate}
		\item Choose a random $r\in \mathbb{Z}^*_n$.
		\item The ciphertext is given by $c=g^m\cdot r^n\;\text{mod}\;n^2$, where $m\in \mathbb{Z}_n, c\in \mathbb{Z}^*_{n^2}$.
	\end{enumerate}
	\item Decryption ($m=\mathcal{D}(c)$):
	\begin{enumerate}
		\item Define the integer division function $L(u) = \frac{u-1}{n}$.
		\item The plaintext is $m=L(c^\lambda\;\text{mod}\;n^2)\cdot \mu\;\text{mod}\;n$.
	\end{enumerate}
\end{itemize}
A cryptosystem is homomorphic if it allows certain computations to
be carried out on the encrypted ciphertext.  The Paillier
cryptosystem is additive homomorphic because the ciphertext of
$m_1+m_2$, i.e., $\mathcal{E}(m_1+m_2)$, can be obtained from
$\mathcal{E}(m_1)$ and $\mathcal{E}(m_2)$ directly:
\begin{equation}
\label{eq:add_explicity}
\begin{aligned}
\mathcal{E}(m_1,r_1)\cdot \mathcal{E}(m_2,r_2) =& (g^{m_1} {r_1}^n)\cdot (g^{m_2} {r_2}^n)\;\text{mod}\;n^2\\
=&(g^{m_1+m_2}(r_1r_2)^n)\;\text{mod}\;n^2\\
=&\mathcal{E}(m_1+m_2, r_1r_2)
\end{aligned}
\end{equation}
The dependency on random numbers $r_1$ and $r_2$ is explicitly shown
in (\ref{eq:add_explicity}),  yet they play no role in the
decryption. For the sake of readability, the following shorthand notation will be used
instead:
\begin{equation}
\label{eq:add}
\mathcal{E}(m_1)\cdot \mathcal{E}(m_2)=\mathcal{E}(m_1+m_2)
\end{equation}
Moreover, if we multiply the same ciphertext $k\in\mathbb{Z}^+$ times, we can obtain
\begin{equation}
\label{eq:mult}
\mathcal{E}(m)^k = \prod_{i=1}^k \mathcal{E}(m)=\mathcal{E}(\sum_{i=1}^km)=\mathcal{E}(km)
\end{equation}
Notice however, the Paillier cryptosystem is not multiplicative
homomorphic because $k$ in (\ref{eq:mult}) is in the plaintext form.
Furthermore, the existence of the random number $r$ in Paillier
cryptosystem gives it resistance to dictionary attacks
\cite{Goldreich_2} which infer a key to an encrypted message by
systematically trying all possibilities, like exhausting all words
in a dictionary. Moreover, since Paillier cryptography only works on
numbers that can be represented by binary strings, we multiply a
real-valued state by a large integer $N$ before converting it to a
binary string so as to ensure small quantization errors. The details
will be discussed in Sec. \ref{sec:quantization}.

\section{Confidential Interaction Protocol}\label{sec:proto}
In this section, we propose a  completely decentralized, third-party
free confidential interaction protocol that can guarantee  average
consensus while protecting the privacy of all participating nodes.
Instead of adding noise to hide the states, our approach combines
encryption with randomness in the system dynamics, i.e., the
coupling weights $a^{(t)}_{ij}$, to prevent two communicating
parties in a pairwise interaction from disclosing information  to each
other. In this way the states are free from being contaminated by
covering noise, guaranteeing a deterministic convergence to the
exact average value.

In this section we present details of our confidential interaction
protocol based on (\ref{eq:ct}) and (\ref{eq:dt}).  In particular we
show how a node can obtain the weighted difference (\ref{eq:diff})
between itself and any of its neighbor  without disclosing each
other's state information:
\begin{equation}
\begin{aligned}
\label{eq:diff}
\Delta x_{ij} =& a_{ij}^{(t)}\cdot(x_j-x_i)\\
\Delta x_{ji} =& a_{ji}^{(t)}\cdot(x_i-x_j)\\
\text{subject to} & \quad a_{ij}^{(t)} = a_{ji}^{(t)}>0
\end{aligned}
\end{equation}
Plugging the state difference (\ref{eq:diff}) into (\ref{eq:ct})
gives a new formulation of  continuous-time average consensus
\begin{equation}
\label{eq:ex_ct}
\dot{x}_i(t) = \sum_{v_j\in N_i} \Delta x_{ij}(t)
\end{equation}
Similarly, we can rewrite the discrete-time consensus update rule as
\begin{equation}
\label{eq:ex_dt}
x_i[k+1] = x_i[k] + \varepsilon\sum_{v_j\in N_i} \Delta x_{ij}[k]
\end{equation}
Notice that in a decentralized system it is impossible to protect
the privacy of both nodes in a pairwise interaction if the protocol
(\ref{eq:diff}) is used without a third party distributing secret
$a^{(t)}_{ij}$. This is due to the fact that even if we encrypt all
the intermediate steps, if one node, for instance $v_i$, has access
to $a^{(t)}_{ij}$, it can still infer the value of $x_j$ through
$x_j = \frac{\Delta x_{ij}}{a^{(t)}_{ij}}+x_i$. From now on, for the
sake of simplicity in bookkeeping, we omit the superscript $t$  in
$a_{ij}^{(t)}$. But it is worth noting that all the results hold for
time-varying weights.

\begin{figure}[t!]
	\centering
	\includegraphics[width=.5\textwidth]{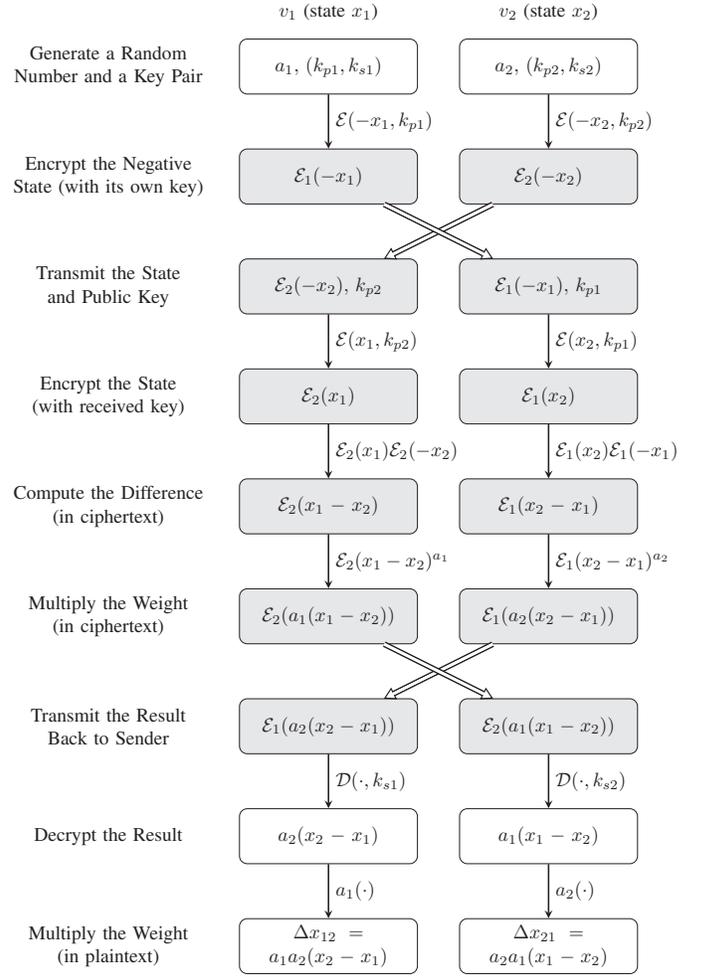}
	\caption{A step-by-step illustration of the confidential interaction protocol. Single arrows indicate the flow of steps; double arrows indicate data exchange via a communication channel. Shaded nodes indicate the computation done in ciphertext. Note that $a_1$ and $a_2$ are different from step to step.}
	\label{fig:1}
\end{figure}

We solve this problem by constructing each weight $a_{ij}$ as the
product of two random numbers, namely $a_{ij}=a_i\cdot a_j=a_{ji}$,
with $0\leq a_i\leq \bar{a}$ (resp. $0\leq a_j\leq \bar{a}$) generated by and only known to node $v_i$ (resp. $v_j$, here $\bar{a}$ is a positive value
denoting the range in implementations which will be explained in
detail later). We will show later that this weight construction
approach renders two interacting nodes unable to infer each other's
state while guaranteeing convergence to the average. Next,
without loss of generality, we use a pair of connected nodes
($v_1,\,v_2$) to illustrate the idea (cf. Fig. \ref{fig:1}). For
simplicity, we assume that the states $x_1$ and $x_2$ are scalar.
Each node maintains its own public and private key pairs $(k_{pi},
k_{si}),\; i\in\{1,\,2\}$.

Due to symmetry, we only show how node $v_1$ obtains the  weighted
state difference $\Delta x_{12}$, i.e., the flow $v_1\rightarrow v_2\rightarrow
v_1$. Before starting the information exchange, node $v_1$ (resp.
$v_2$) generates its new non-negative random number $a_1$ (resp.
$a_2$) which is  within a certain range $[0,\,\bar{a}]$ in
implementation. First, node $v_1$ sends its encrypted negative state
$\mathcal{E}_1(-x_1)$ as well as the public key $k_{p1}$ to node
$v_2$. Note that here the subscript in $\mathcal{E}_1$ denotes
encryption using the public key of node $v_1$. Node $v_2$ then
computes the encrypted $a_2$-weighted difference
$\mathcal{E}_1\left(a_2(x_2-x_1)\right)$ following the three steps
below:
\begin{enumerate}
	\item Encrypt $x_2$ with $v_1$'s public key $k_{p1}$: $x_2 \rightarrow \mathcal{E}_1(x_2)$.
	\item Compute the difference directly in ciphertext:
	\begin{equation}
	\mathcal{E}_1(x_2-x_1)=\mathcal{E}_1(x_2+(-x_1))=\mathcal{E}_1(x_2)\cdot \mathcal{E}_1(-x_1)
	\end{equation}
	\item Compute the $a_2$-weighted difference in ciphertext:
	\begin{equation}
	\label{eq:exp_a1}
	\mathcal{E}_1\left(a_2 (x_2-x_1)\right)=\left(\mathcal{E}_1(x_2-x_1)\right)^{a_2}
	\end{equation}
\end{enumerate}
Then $v_2$ returns $\mathcal{E}_1\left(a_2 (x_2-x_1)\right)$ to
$v_1$. After receiving $\mathcal{E}_1\left(a_2
(x_2-x_1)\right)$, $v_1$ decrypts it using the private key $k_{s1}$
and multiplies the result with $a_1$ to get the  weighted difference
$\Delta x_{12}$:
\begin{equation}
\begin{aligned}
\label{eq:mul_a1}
\mathcal{E}_1\left(a_2(x_2-x_1)\right)\xrightarrow{\mathcal{D}_1}a_2(x_2-x_1)\\
\Delta x_{12}=a_1a_2(x_2-x_1)
\end{aligned}
\end{equation}

In a similar manner, the exchange $v_2\rightarrow v_1 \rightarrow v_2$ produces $\mathcal{E}_2\left(a_1(x_1-x_2)\right)$ for $v_2$ who then decrypts the message and multiplies the result by its own multiplier $a_2$ to get $\Delta x_{21}$
\begin{equation}
\begin{aligned}
\label{eq:mul_a2}
\mathcal{E}_2\left(a_1(x_1-x_2)\right)\xrightarrow{\mathcal{D}_2}a_1(x_1-x_2)\\
\Delta x_{21}=a_2a_1(x_1-x_2)
\end{aligned}
\end{equation}
After each node collects the weighted differences from all
neighbors, it updates its state with (\ref{eq:ct}) or (\ref{eq:dt})
accordingly.

Several remarks are in order:
\begin{itemize}
	\item The  construction of each $a_{ij}$ as the product of two
	random
	numbers $a_i$ and $a_j$ is key to guarantee that the weights are
	symmetric, i.e., $a_{ij}=a_{ji}$, which is crucial for average consensus \cite{Olfati-Saber2007}.
	\item $v_2$ does not have the private key of $v_1$ and cannot see $x_1$ which is encrypted in $\mathcal{E}_1(-x_1)$.
	\item Given $a_2(x_2-x_1)$, $v_1$ cannot solve for $x_2$ because $a_2$ is only known to $v_2$.
	\item At each iteration, real-valued states are converted to fixed point representation for encryption; the weighted differences are converted back to real values for update.
	\item We encrypt $\mathcal{E}_1(-x_1)$ because it is more difficult to compute subtraction in ciphertext. The issue regarding encrypting negative values using Paillier is discussed in Sec. \ref{sec:impl}.
\end{itemize}

\section{Theoretical Analysis of Convergence}\label{sec:analysis}

In this section,  we discuss theoretically the convergence property
under the confidential interaction protocol and   how to set the
appropriate values for the multiplier $a_i$ (and $\varepsilon$ for
DT).

\subsection{Convergence for Continuous-Time Consensus}
Let $\mathbf{x}\in \mathbb{R}^M$ denote the augmented state vector
of all nodes. The network dynamics in (\ref{eq:ct}) can be rewritten
as:
\begin{equation}
\dot{\mathbf{x}} = -\mathbf{L}^{(t)} \mathbf{x}(t)
\end{equation}
where $\mathbf{L}^{(t)}=[l^{(t)}_{ij}]$ is the time-varying Laplacian matrix defined by
\begin{equation}
\begin{aligned}
\label{eq:structure}
l^{(t)}_{ij}=&\begin{cases}
\sum_{v_j\in N_i} a^{(t)}_{ij} & i= j
\\    -a^{(t)}_{ij} & i \neq j
\end{cases}
\end{aligned}
\end{equation}

\begin{Theorem 1}
	If the coupling weights $a^{(t)}_{ij}$ in  (\ref{eq:structure}) are established according to the confidential interaction protocol in Sec. \ref{sec:proto}, then under any positive bound
	$\bar{a}>0$,
	the system will achieve average consensus  with states
	converging
	to
	\begin{equation}
	\lim_{t\to\infty} \mathbf{x}(t) = \alpha\mathbf{1}\quad\text{where}\;\alpha = \text{Avg}(0)=\frac{1}{M}\mathbf{1}^T \mathbf{x}(0)
	\end{equation}
\end{Theorem 1}

\noindent \textit{Proof}: It is already known that average consensus
can be achieved if for all time $t_0>0$, there exists a constant
$\bar{T}>0$ such that
$a_{ij}^{(t)}> 0$ is true for some $t\in[t_0,\,t_0+\bar{T}]$ \cite{Olfati-Saber2007,Moreau05}. Noting that the weights $a_{ij}^{(t)}\geq0$ obtained from the
confidential interaction protocol in Sec. \ref{sec:proto}
are random and independent of each other, the proof can be obtained by following the line of reasoning in
\cite{Olfati-Saber2007,Moreau05}. \hfill{$\blacksquare$}

\subsection{Convergence for Discrete-Time Consensus}

In discrete-time domain (\ref{eq:dt}) can be rewritten as
\begin{equation}
\mathbf{x}[k+1]=\mathbf{P}^{(k)}\mathbf{x}[k]
\label{eq:perron}
\end{equation}
where $\mathbf{P}^{(k)}=\mathbf{I}-\varepsilon \mathbf{L}^{(k)}$ is the Perron matrix and $\mathbf{L}^{(k)}=[l^{(k)}_{ij}]$ is the time-varying Laplacian matrix defined by
\begin{equation}\label{eq:weithts in DT}
\begin{aligned}
l^{(k)}_{ij}=&\begin{cases}
\sum_{v_j\in N_i} a^{(k)}_{ij} & i= j
\\    -a^{(k)}_{ij} & i \neq j
\end{cases}
\end{aligned}
\end{equation}

\begin{Theorem 2}
	If the coupling weights $a^{(k)}_{ij}$ in  (\ref{eq:weithts in DT})
	are established according to the confidential interaction protocol
	in Sec. \ref{sec:proto} and $\varepsilon$ satisfies
	$0<\varepsilon<\frac{1}{\Delta}$ where $\Delta=\max_{i}|N_i|$ with
	$|\bullet|$ denoting the set cardinality, then under any positive
	bound $0<\bar{a}<1$,
	the system will achieve average consensus  with states
	converging
	to
	\begin{equation}
	\lim_{k\to\infty}\mathbf{x}[k] = \alpha \mathbf{1}\quad\text{with }\alpha=\text{Avg}[0]=\frac{1}{M}\mathbf{1}^T\mathbf{x}[0]
	\end{equation}
\end{Theorem 2}
\textit{Proof}: The proof can be obtained by following the similar line of reasoning of Corollary 2 in \cite{Olfati-Saber2007}.
\hfill{$\blacksquare$}

\begin{Remark 1}
	Since the framework allows time-varying weighted adjacency matrix $\mathbf{A}^{(t)}$ for the discrete-time domain
	or $\mathbf{A}^{(k)}$ for the discrete-time domain, it can  easily be  extended  to the case with switching interaction graphs
	according to   \cite{Moreau05}.
\end{Remark 1}

\section{Analysis of Privacy and Security}\label{sec:security}
Privacy and security are often used interchangeably in the
literature but here we make the distinction explicit.  Among the
control community privacy is equivalent to the concept of
unobservability. Privacy is also closely related to the concept of
semantic security from cryptography \cite{Goldreich_2}. Both
concepts essentially concern with an honest-but-curious adversary
which is interested in learning the states of the network but
conforms to the rules of the system. Security, on the other hand,
deals with a broader issue which includes learning the states as
well as the possibilities of exploiting the system to cause damages.

\subsection{Privacy Guarantees}
Our protocol provides protection against an honest-but-curious
adversary or an observer
eavesdropping the communication. In the literature, an honest-but-curious
adversary is usually defined as a node who
follows all protocol steps correctly but is curious and collects
received  data in an attempt to learn some information about other
participating parties. An observer eavesdropping the communication is usually defined as an adversary who is able to intercept exchanged messages and read the bits within. In this paper, for the sake of simplicity, we generally refer to both types of adversaries as honest-but-curious adversaries.   

The Paillier encryption algorithm is known to provide semantic
security,  i.e., Indistinguishability under Chosen Plaintext Attack
(IND-CPA) \cite{Paillier1999}. As a result, the recipient of the
first transmission $\mathcal{E}_i(-x_i)$ cannot see the value of $x_i$
at any time. We now prove that an honest-but-curious adversary
cannot infer the initial state of a neighbor even if it can
accumulate and correlate the return messages $a_ia_j(x_j-x_i)$ in
multiple steps (except in a trivial case that should always be
avoided, as explained in Theorem \ref{thm:4}).

As per the naming convention in cryptography, it is customary to
name the legitimate sender and receiver participants as $A$ (Alice)
and $B$ (Bob), and the adversary as $E$ (Eve).

\begin{figure}[t]
	\centering
	\includegraphics[width=0.35\textwidth]{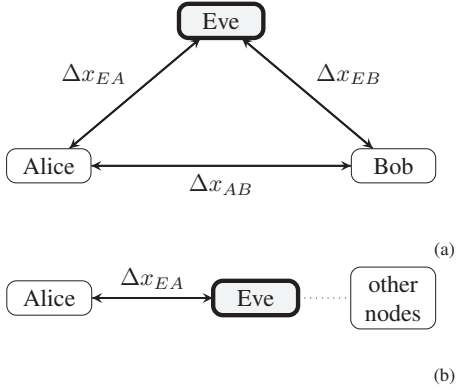}
	\caption{A node should be connected to at least one legitimate node  to prevent privacy leak.}\label{fig:privacy}
\end{figure}

\begin{Theorem 3}\label{thm:3}
	Assume that all nodes follow the confidential interaction  protocol.
	An honest-but-curious node Eve cannot learn the initial state of a neighboring node Alice if Alice is also
	connected to another legitimate node Bob.
\end{Theorem 3}

\textit{Proof}: Without loss of generality, we consider the connection configuration illustrated in Fig. \ref{fig:privacy} (a) where Eve can
interact with   both Alice and Bob. If Eve cannot infer the state of
Alice or Bob in this
configuration, neither can it  when either the Alice---Eve connection or the Bob---Eve connection is removed which reduces the information accessible to Eve.

From the perspective of the honest-but-curious node Eve, the measurements seen at each time step $k$ are
$\Delta x_{Ei}[k]=a^{(k)}_ia^{(k)}_E(x_i[k]-x_E[k]),\,i\in\{A, B\}$. In matrix form, define these observations as $y_E[k]$
\[
y_E[k]=[\Delta x_{EA}[k], \ \Delta x_{EB}[k]]^T=\mathbf{C}^{(k)}_E \mathbf{x}[k]
\] where:
\begin{align}
	\mathbf{C}^{(k)}_E= \begin{bmatrix}
		-a^{(k)}_Aa^{(k)}_E & a^{(k)}_Aa^{(k)}_E& 0\\
		-a^{(k)}_Ba^{(k)}_E & 0 & a^{(k)}_Ba^{(k)}_E
	\end{bmatrix}
\end{align}
It can be easily derived that after $K$ steps, the measurements
collected by Eve are given by

\begin{equation}    \label{eq:obsr}
\mathbf{y}_E[0:K] = \mathbf{O}_{E,[0:K]}\mathbf{x}[0]
\end{equation}
where the observability matrix $\mathbf{O}_{E,[0:K]}$ is given by
\begin{equation}\label{eq:observability matrix}
\mathbf{O}_{E,[0:K]} = \begin{bmatrix}
\mathbf{C}_E^{(0)}\\ \mathbf{C}_E^{(1)}\mathbf{P}^{(0)} \\ \vdots \\ \mathbf{C}_E^{(K)}\prod_{k=K-1}^{0}\mathbf{P}^{(k)}
\end{bmatrix}
\end{equation}
with $\mathbf{P}^{(k)}$ being the Perron matrix defined in
(\ref{eq:perron}).
Note that   the
entries of $\mathbf{C}^{(k)}_E$ and $\mathbf{P}^{(k)}$ are unknown to Eve
because $a^{(k)}_A$ and $a^{(k)}_B$ are randomly
chosen by Alice and Bob respectively. Therefore, the ability for Eve to infer
the state of other nodes \emph{cannot} be analyzed using
conventional observability based approach in e.g.,
\cite{observability1,observability2}.

We propose a new analysis approach based on the solvability of systems of equations.
From (\ref{eq:obsr}) it can be seen that  Eve can establish
$2(K+1)$  equations based on received information from time instant
0 to $K$. Given that after consensus, Eve can know the final state
of other nodes which is equal to its own final state (represent it
as $\alpha_{\rm consensus}$),  it can establish one more equation
\begin{equation}\label{eq:additional_equation}
x_A[0]+x_B[0]+x_E[0]=3\alpha_{\rm consensus},
\end{equation}
which makes the number of   employable equations   to $2(K+1)+1$. If
there are more than $2(K+1)+1$ unknowns involved in these
$2(K+1)+1$ equations, then it is safe to say that Eve cannot solve
the equations and get the initial states of $x_A[0]$ and $x_B[0]$.
In fact, the confidential interaction protocol introduces $2(K+1)$
unknown parameters  $a_A^{(0)},\, a_A^{(1)},\,\cdots,a_A^{(K)},\,a_B^{(0)},\,
a_B^{(1)},\,\cdots,a_B^{(K)}$, which, in combination with
$x_A[0], x_B[0]$ unknown to Eve, will make the total number of unknowns to
$2(K+1)+2$.  Therefore, the honest-but-curious Eve
cannot use the accessible $2(K+1)+1$ system of  equations in (\ref{eq:obsr})
to solve for the initial states of  $x_A[0]$ and $x_B[0]$. \hfill{$\blacksquare$}

\begin{Remark 2}
	Following the same line of reasoning, it can be obtained that an
	honest-but-curious node Eve 1 cannot infer the initial state of a
	neighboring node Alice if Alice is also connected to another
	honest-but-curious node Eve 2 that does not collude with Eve 1.
\end{Remark 2}

Based on the analysis framework, we can also obtain a situation in
which it is possible for Eve to infer other nodes' states which
should be avoided.

\begin{Theorem 4}\label{thm:4}
	If a node Alice is connected to the rest of the network only through an
	(or a group of colluding) honest-but-curious node(s) Eve, then Alice's
	initial state can be inferred by Eve.
\end{Theorem 4}

\textit{Proof}: If Alice is directly connected to multiple
honest-but-curious   nodes that collude with each other, then these
nodes can  share information with each other to cooperatively estimate
Alice's state, and hence can be regarded as one node. Therefore,
we just consider the case where Alice is only
connected to one   honest-but-curious node Eve, as illustrated in  Fig. \ref{fig:privacy}
(b).
In this case, from the perspective of the honest-but-curious node Eve, the measurement seen at each time step $k$
is
$\Delta x_{EA}[k]=a^{(k)}_Aa^{(k)}_E(x_A[k]-x_E[k])$.
Similar to the proof of Theorem \ref{thm:3}, we can write the measurements accessible to Eve in
a
matrix form
$y_E[k]=[\Delta x_{EA}]=\mathbf{C}^{(k)}_E \mathbf{x}[k]$, where
\begin{align}
	\mathbf{C}^{(k)}_E= \begin{bmatrix}
		-a^{(k)}_Aa^{(k)}_E & a^{(k)}_Aa^{(k)}_E\end{bmatrix}
\end{align}
After $K$ steps, the measurements collected by Eve are given by
\begin{equation}    \label{eq:obsr2}
\mathbf{y}_E[0:K] = \mathbf{O}_{E,[0:K]}\mathbf{x}[0]
\end{equation}
with
the observability matrix $\mathbf{O}_{E,[0:K]}$ having the same form
as (\ref{eq:observability matrix}).

Now in the $K+1$ equations collected by Eve in (\ref{eq:obsr2}),
there are $K+2$ unknowns  $x_A[0],\, a_A^{(0)},\, a_A^{(1)},\,\cdots,a_A^{(K)}$.
However, after converging to average consensus, Eve will be able to
know the final state of other nodes (the same as its final state), which enables it to
construct another equation about the initial states like
(\ref{eq:additional_equation}). This will make the total number
of equations equal to the total number of involved unknowns and
make solving initial state of $x_A$ possible.
\hfill{$\blacksquare$}

Next we use an example to illustrate that it is indeed possible for
Eve to infer the state of Alice if Eve is Alice's only neighbor.
Consider the configuration in Fig. \ref{fig:privacy} (b). Eve
receives $\Delta x_{EA}[k]=-\Delta x_{AE}[k]$ from Alice. In
addition, when the protocol converges after \textit{K} steps, Eve
knows the final state which is identical for all the nodes. The
initial value of Alice can be simply inferred by Eve through
\begin{equation}
\begin{aligned}
x_A[0] &= x_A[K] + \varepsilon \sum_{k=0}^{K-1}\Delta x_{AE}[k]\\
&= x_E[K] - \varepsilon \sum_{k=0}^{K-1}\Delta x_{EA}[k]
\end{aligned}
\end{equation}
Therefore this single connection configuration should always be
avoided,  which is also required by other data-obfuscation based privacy
protocols, for instance in  \cite{Manitara13} and \cite{Mo17}.

\subsection{Security Solution}
Due to the additive homomorphic property, the Paillier cryptosystem
is vulnerable to active adversaries who are able to alter the
message being sent through the channel. Although such adversaries
cannot  find out the exact states of the communicating nodes, they
can still inflict significant damage to the system.

\begin{figure}[t]
	\centering
	\includegraphics[width=0.4\textwidth]{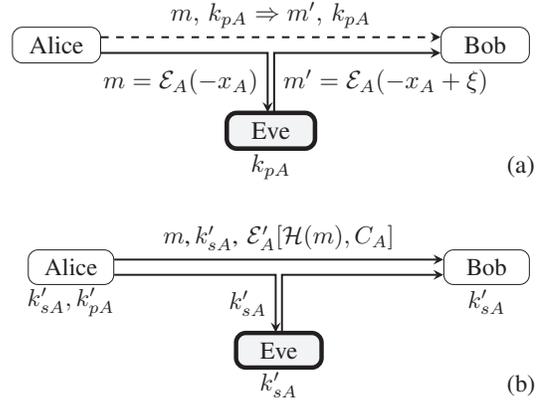}
	\caption{Illustration of attacks from an active attacker (a) and the defense mechanism  with a digital signature (b).}\label{fig:security}
\end{figure}

Consider the scenario where the communication from node Alice to Bob
is intercepted by an active adversary Eve (cf. Fig.
\ref{fig:security} (a)). Since Alice's public key $k_{pA}$ is sent
along with $\mathcal{E}_A(-x_A)$, Eve may use the additive
homomorphism to inject an arbitrary noise $\xi$ to    the original
message $\mathcal{E}_A(-x_A)$ to sway it to $\mathcal{E}_A(-x_A+\xi)$.
If Bob has no way to tell if the received message has been modified,
Eve may exploit this vulnerability to make the network either
converge to a wrong value or not converge at all.

In applications where security is of prime concern,
it is imperative to be able to verify the  integrity of any incoming message. We
propose to attach a digital signature to the exchanged message in
the confidential interaction protocol, based on which the recipient
can verify possible modifications during
communication. The signature requires an additional pair of
public/private keys ($k'_{pA},k'_{sA}$) and a hash function
$\mathcal{H}(\cdot)$, and is represented as ($k_{sA}'$,
$\mathcal{E}'_A\left[\mathcal{H}(m), C_A\right])$, where $C_A$ is an  unforgeable certificate assigned by an authority. The additional private
key $k'_{sA}$ is sent so that Bob can decrypt
$\mathcal{E}'_A\left[\mathcal{H}(m)\right]$ and check if the
resulting $\mathcal{H}(m)$ matches the received $m$ in terms of the   hash operation $\mathcal{H}(\cdot)$ (cf. Fig. \ref{fig:security} (b)).
Because without the public key $k'_{pA}$, Eve cannot forge a valid
signature (that can be decrypted by Bob), any Eve's attempt to
modify $m$ will cause a mismatch between received $m$ and decrypted
$\mathcal{H}(m)$ in terms of the   hash operation
$\mathcal{H}(\cdot)$.

\section{Implementation Details}\label{sec:impl}
In addition to the constraints imposed on $a_i$ and $\varepsilon$,
there are other technical issues that must be addressed  for the
implementation of our confidential interaction protocol.

\subsection{Quantization}\label{sec:quantization}
Real-world applications typically have $x_i\in \mathbb{R}$ which are
represented by floating point numbers in modern computing
architectures. On the contrary, encryption algorithms only work on
unsigned integers. Define the casting function $f(\cdot, \cdot):
\mathbb{R}\times \mathbb{R}\rightarrow \mathcal{M}\subset
\mathbb{Z}$ and its inverse $f^{-1}(\cdot, \cdot):\mathcal{M}\times
\mathbb{R}\rightarrow \mathbb{R}$ as
\begin{equation}
\begin{aligned}
&f(x, N) = \left\lceil Nx \right\rfloor_\mathcal{M},\quad
&f^{-1}(y, N) = \frac{y}{N}
\end{aligned}
\end{equation}
where $\left\lceil \cdot \right\rfloor_\mathcal{M}$ maps the input
to the nearest integer in $\mathcal{M}$. For the Paillier
cryptosystem, this mapping is equivalent to the rounding operation,
hence the step size is $\Delta Y=1$ which is uniform. Consequently the
maximum quantization error is bounded by
\begin{equation}
\begin{aligned}
\max_{x\in \mathbb{R}}|x-f^{-1}(f(x, N),N)|= \frac{\Delta Y}{N}
\end{aligned}
\end{equation}

In practice we choose a sufficiently large value for  $N$ so that
the quantization error is negligible. This is exactly how we convert
the state $x_i$ of a node from real value to a fixed length integer
and back to a floating point number. The conversion is performed at
each iteration of the protocol.


\subsection{Subtraction and Negative Values}
Another issue is how to treat the sign of an integer for encryption.
\cite{homomorphic_privacy1} solves this problem by mapping negative
values to the end of the group $\mathbb{Z}_n$ where $n=pq$ is given
by the public key. We offer an alternative solution by taking
advantages of the fact that encryption algorithms blindly treat bit
strings as unsigned integers. In our implementation all integer
values are stored in fix-length integers (i.e., long int in C) and
negative values are left in two's complement format. Encryption and
intermediate computations are carried out as if the underlying data
were unsigned. When the final message is decrypted, the overflown
bits (bits outside the fixed length) are discarded and the remaining
binary number is treated as a signed integer which is later
converted back to a real value. 

\subsection{Implementation on Raspberry Pi}\label{sec:example}
To confirm the effectiveness of the secure and privacy-preserving
average consensus approach in real-world cyber-physical systems, we
implemented the algorithm on six Raspberry Pi boards with
64-bit ARMv8 CPU and 1 GB RAM.

\begin{figure}[tpb]
	\centering
	\includegraphics[width=0.43\textwidth, trim=.5cm .0cm .5cm .5cm,clip]{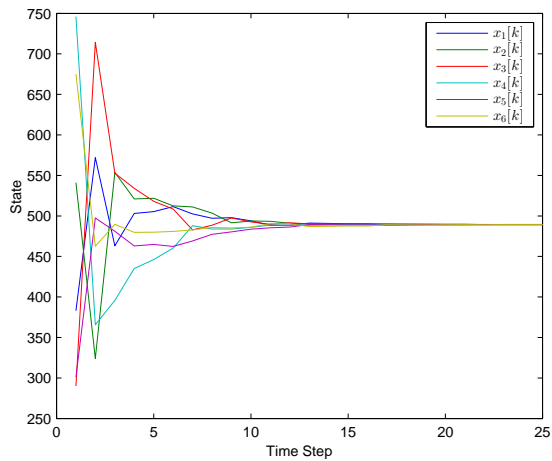}
	\caption{All nodes converge to the  average consensus value in the experimental verification using Raspberry Pi boards. The states have initial values as 290, 746, 541, 383, 301, and 675, respectively and they all  converge to the average consensus value 489.33 in about 13 steps.}
	\label{fig:experimental_plot}
\end{figure}

In the implementation, the communication was conducted through Wi-Fi
based on the ``sys/socket.h"  C library.   Paillier encrption and decryption  were realized using  the ``libpaillier-0.8"
library from \cite{Paillier_lib}. To obtain $\Delta x_{ij}$ in a
pair-wise interaction, a node employs a request message to
initialize the interaction and the other node replies with a
response message.  In a multi-node network, for a node to be able to
simultaneously receive requests and responses from multiple
neighbors, parallelism needed to be introduced.
The ``pthread" C library was used to generate multiple parallel
threads to handle incoming requests and responses. Each time a node receives a request/response, it generates
a new thread to handle it and immediately listens for more requests.
Because in the implementation, it is impossible to start all nodes
simultaneously, a counter is introduced on each node and its value is embedded
in each request/response packet to help  nodes make sure that they are on the same pace. For 64 byte encryption key, the
size of the actual packet is 144 bytes, which includes all necessary
headers and stuffing bytes. For each interaction, the average
processing latency was 7.8 ms, which is acceptable for most real-time cyber-physical systems. The implementation result is given in
Fig. \ref{fig:experimental_plot}, which shows that perfect consensus can be achieved.

\section{Conclusions}\label{sec:conclusion}
In this paper we proposed a decentralized secure and
privacy-preserving protocol for the network average consensus problem. In
contrast to previous approaches where the states are covered with
random noise which unavoidably affects the convergence performance,
we encode randomness to the system dynamics with the help of an
additive homomorphic cryptosystem which allows the convergence to
the exact average in a deterministic manner. The protocol also
 allows easy incorporation of active attacker defending mechanisms. 
 Although   
our approach has higher computational complexity compared to the unencrypted
alternatives, experimental results on Raspberry Pi  confirm that the  computational burden is   manageable on resource-restricted cyber-physical systems.

\section*{Acknowledgement}
The authors would like to thank Christoforos Hadjicostis and Yilin
Mo for their   comments on an initial draft of this article. This work was supported in part by the National Science Foundation
under Grant 1738902.

\bibliographystyle{ACM-Reference-Format}
\bibliography{reference1} 

\end{document}